\documentclass{llncs}

\usepackage{graphicx}
\usepackage{url}

\begin{document}

\title{ROMEO: ReputatiOn Model Enhancing \\OpenID Simulator}

\author{Gin\'es D\'olera Tormo\inst{1} \and F\'elix G\'omez M\'armol\inst{1} \and Gregorio Mart\'inez P\'erez\inst{2}}
\institute{NEC Europe Ltd., Kurf\"ursten-Anlage 36, 69115 Heidelberg, Germany \\
\email{\{gines.dolera|felix.gomez-marmol\}@neclab.eu}
\and
Department of Information and Communications Engineering, University of Murcia, Murcia, 30100 Spain \\
\email{gregorio@um.es}}

\maketitle

\begin{abstract}

OpenID is a standard decentralized initiative aimed at allowing Internet users to use the same personal account to access different services. Since it does not rely on any central authority, it is hard for such users or other entities to validate the trust level of each entity deployed in the system. Some research has been conducted to handle this issue, defining a reputation framework to determine the trust level of a relying party based on past experiences. However, this framework has been proposed in a theoretical way and some deeper analysis and validation is still missing. Our main contribution in this paper consist of a simulation environment able to validate the feasibility of the reputation framework and analyze its behaviour within different scenarios.

\end{abstract}

\section{Introduction}

OpenID~\cite{idm:2006:recordon:2006:dim} is an open technology standard which defines a decentralized authentication protocol in order to allow end-users to sign in to multiple websites with the same user account. It allows users to maintain their private information in a single point, and release it to external entities when it is required in a controlled way. For example, users could reveal their credit card number, which is securely stored in their OpenID providers, to purchase an item in an online shop.

Due to its decentralized nature, OpenID technology does not rely on any central authority which validates the trust level of the entities involved in the system, such as the different OpenID providers or the service providers. In this way, it is hard for users to decide whether a given service provider is trustworthy enough to share their private information before interacting with it, which actually implies sending such private information.

A reputation framework to be integrated with OpenID was defined in \cite{idm-trm:2011:gomez-marmol:atc}, without requiring any central authority intervention. It describes how the OpenID protocol can be enhanced so that each OpenID provider can collect recommendations from different users on a given relying party, even if they belong to different OpenID providers. These recommendations can be aggregated appropriately to provide useful information about the service to the users, such as whether it is trustworthy or not, before they start interacting with it. Such reputation framework, however, is described in a theoretical way and there is not a reference implementation or detailed description of the internal component which can prove the feasibility of such a framework. 


Our main contribution in this paper is presenting a simulation environment, developed within NEC Laboratories Europe, able to analyze and validate the feasibility of a number of reputation engines integrated with OpenID, including the ones described in \cite{idm-trm:2011:gomez-marmol:atc}. This simulator environment, entitled ROMEO (ReputatiOn Model Enhancing OpenID Simulator), allows evaluating the capability of malicious users or entities to exploit the reputation system. For instance, it is able to analyze whether a relying party could increase its reputation introducing biased recommendations to the system~\cite{trt:2009:gomez-marmol:cose}.

The remainder of the article is organized as follows. Section~\ref{sec:background} presents some related work to motivate our contribution. Section~\ref{sec:features} introduces threats which any simulator environment should consider to properly analyze this kind of systems. Section~\ref{sec:architecture} describes the internal components which define the architecture of the framework, while section~\ref{sec:romeo} presents the user interface of the simulator.

\section{Background}\label{sec:background}

The reputation framework described in~\cite{idm-trm:2011:gomez-marmol:atc} was designed for enhancing the OpenID users' experience when accessing a relying party. This framework allows OpenID providers to aggregate recommendations about a given relying party from users which have already interacted with it. This allows OpenID providers to inform other users before interacting with the given relying party. 

Since some of the OpenID providers would not have enough users to form accurate reputation scores, that framework describes a mechanism to gather recommendations from external sources. To know where to collect the recommendations from, the OpenID providers ask the relying party for potential recommenders. The recommenders are mainly other OpenID providers which have recently interacted with the relying party, hence having updated recommendations of the provided service.

Additionally, this framework considers users preferences to provide customized reputation values~\cite{trt:2007:ziegler:dss}. When a reputation value is being computed for a given user, the recommendations given by users with similar preferences to the given one will have a higher importance in the computation of such reputation score.

Nevertheless, some of the gathered recommendations could be malicious or inaccurate, for example if they are biased trying to increase or decrease the reputation of certain relying parties. To solve this problem, the reputation framework establishes dynamic weights to the sources of recommendations based on its accuracy. Therefore, inaccurate recommendations sources will be punished, and its recommendations will not be taken into account. However, this framework only has been defined theoretically and a deep analysis and validation is still missing.

Several reputation simulators have been proposed to validate the feasibility of reputation frameworks. ART testbed~\cite{trm-sim:2006:fullan:itrust} and TOSim~\cite{trm-sim:2007:zhang:iccs} have become a reference due to its flexibility and modularity, allowing researchers to easily compare and analyze their trust and reputation models. P2P-SIM~\cite{trm-sim:2010:west:igi-global} presents a simulator for evaluating P2P-based trust and reputation management algorithms. It simulates the network structure, and outputs statistics about how the trust management is performed. In a similar way, TRMSim-WSN~\cite{trm-sim:2009:gomez-marmol:ieee-icc} allows testing and compare trust and reputation models, but within Wireless Sensor Networks (WSN) scenarios. However, we have not found any testbed or simulator targeting reputation management mechanisms integrated within user-centric identity management systems. Those simulation environments should take into account the particularities of these specific reputation models, such as computing reputation based on users' preferences.

\section{OpenID-integrated reputation frameworks threats}\label{sec:features}

There are some aspects which can compromise the behaviour of the presented reputation framework, and should be analyzed to determine its feasibility. In the following, we present some relevant features and threats which should be taken into consideration when designing a simulator. 

\begin{itemize}

\item Relying parties could know the recommendation that each OpenID provider has about them. Even though the framework just share recommendations between OpenID providers, it would be easy for any relying party to deploy an OpenID provider which asks for recommendation to the rest of them.   

\item Relying parties could offer services with diverse qualities. For instance, a relying party could give low quality on the main services but offering several low-cost with high quality services just for increasing its reputation, if it was calculated as a global reputation instead of depending of the service.

\item Quality of the services offered by the relying parties could vary during the time. For example, a relying party with good reputation could suddenly decrease the quality of its services.

\item Relying Parties could fake the list of potential recommenders by including just the recommenders which provide better recommendations about them. In this case, the relying parties could try to increase their reputation since only biased recommendation sources are taken into consideration. 

\item A relying party could decide not to participate in the reputation framework. For example, due to lack of resources or to avoid low reputation values.

\item There could be malicious OpenID providers which supply inaccurate recommendations values trying to either increase or decrease the reputation value of a given relying party.

\item Users could provide inaccurate recommendations. That includes some users who make mistakes when providing recommendations, and those deployed by a malicious relying party trying to increase its reputation~\cite{trt:2011:Borg:ccis}. 

\item The framework cannot assume unlimited resources to perform any of its processes. Some of the defined processes may require a vast amount of storage, computer or network resources to work properly. Furthermore, the response times are critical to determine the success of the reputation framework.

\item A malicious entity can present multiple identities, issuing a higher fraction of the recommendations of the system, as a Sybil attack~\cite{idm:2002:douceur:iptps}.


\end{itemize}

\section{ROMEO Architecture overview}\label{sec:architecture}

\begin{figure}[!h]
   \centering
   \includegraphics[width=0.7\columnwidth]{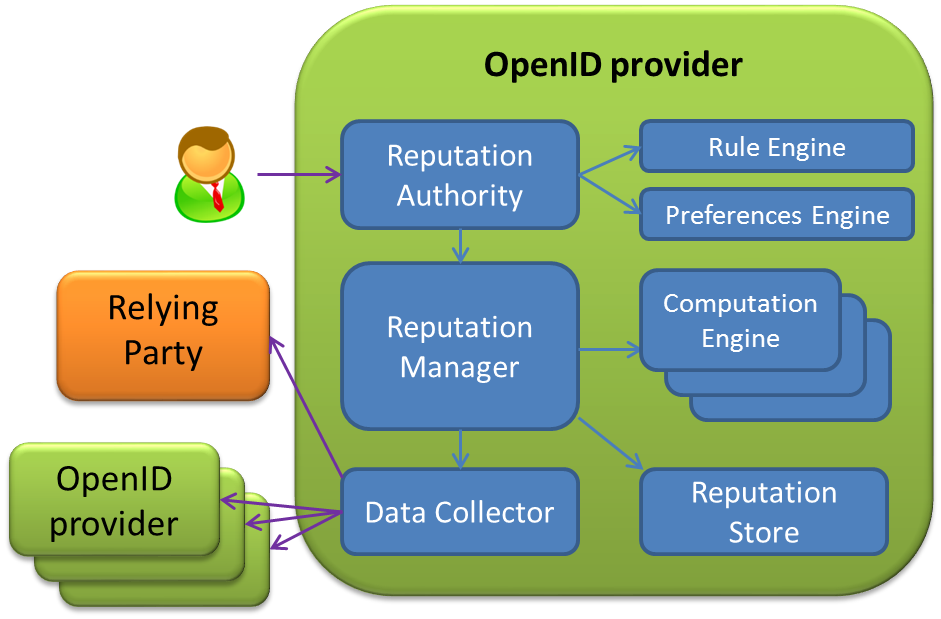}
   \caption{General ROMEO architecture overview}
   \label{fig:architecture}
 \end{figure}

This section presents the internal components of the architecture developed to simulate the reputation framework, as shown in Figure~\ref{fig:architecture}. The architecture has been designed to allow an easy extension in order to validate different reputation computation engine that may be defined in the future as part of OpenID. Next, we describe each of the components belonging to the architecture.

\begin{itemize}

\item \textbf{Reputation Authority}. This component is the interface for external entities to make use of the reputation framework functionalities. Either the OpenID provider in use sends internal queries to this module to obtain the reputation value of a relying party or other OpenID providers ask this component for gathering external recommendations. Additionally, this module is also utilized by the users to provide recommendations once they have interacted with the relying party, which feed the reputation framework.

\item \textbf{Rule Engine}. Computing reputation values could be directed by certain dynamic rules. For instance, if the system is overloaded, just the last 25 recommendations will be taken into account instead of all of them. The Rule Engine component is in charge of evaluating these rules and influence the rest of the reputation computation process. 

\item \textbf{Preferences Engine}. The preferences of the user which is asking for reputation could be taken into account when computing the reputation values. This component processes the preferences of the user and establishes certain parameters according to them, which will influence the weights of the recommendations of other users when computing the reputation score.

\item \textbf{Reputation Manager}. This component is the controller of the reputation architecture. By using the rest of the components, this component manages the internal and external recommendations and aggregates them using a specific Computation Engine when a reputation value is requested.

\item \textbf{Reputation Store}. This component is in charge of maintaining recommendations from different users or OpenID providers which have been gathered in the past. It also maintains a weight together with each recommendation in order to determine the accuracy of the recommender. Additionally, this module may act as a cache by storing computed reputation values during a while, which avoids computing the same reputation value again when the users ask for it in a short period of time.

\item \textbf{Data Collector}. The Data Collector component is used by the Reputation Manager to collect external recommendations. This component retrieves the list of potential recommenders from a given relying party and then asks each of these recommenders (i.e. other OpenID providers) for recommendations about the relying party.

\item \textbf{Computation Engines}. This set of components is the core of the reputation framework since it is in charge of aggregating reputation values from the recommendations and its associated weights. A different Computation Engine could be used depending on some systems conditions, such as the current network load. Hence, a common interface is defined so they could be interchangeable. Additionally, this module also adjusts the weights of the recommendations sources based on their accuracy in order to punish or reward such recommendations sources for future reputation computation.

\end{itemize}

\section{Reputation Model Enhancing OpenID Simulator}\label{sec:romeo}

ROMEO (ReputatiOn Model Enhancing OpenID) is the performed simulator, within NEC Laboratories Europe, aimed to evaluate reputation frameworks integrated with OpenID, such as the one presented in~\cite{idm-trm:2011:gomez-marmol:atc}. It allows evaluating the capability of malicious users or entities (or groups of them) to exploit the reputation system. That is, if they could increase or decrease the reputation of a given relying party by supplying biased recommendations. Furthermore, it analyzes how different reputation computation engines are adapted to different scenarios.

 \begin{figure}[!h]
   \centering
   \includegraphics[width=1.0\columnwidth]{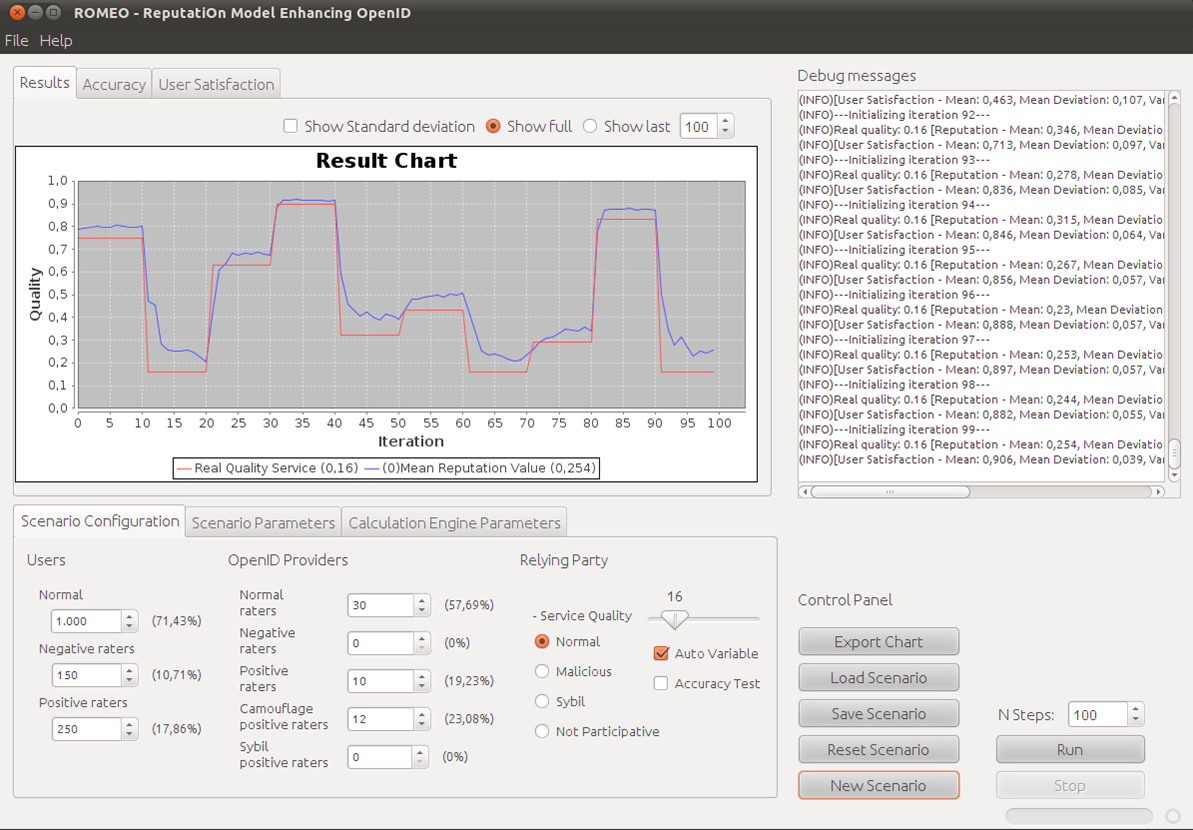}
   \caption{ROMEO simulator graphical interface screenshot}
   \label{fig:romeo}
 \end{figure}%
 
\subsection{Scenario elements}

As shown in Figure~\ref{fig:romeo}, ROMEO presents a graphical interface where different reputation-based scenarios could be defined. A scenario is composed of simulated users belonging to different simulated OpenID providers interacting with a specific relying party under certain conditions. Such conditions are parametrized to configure the behaviour of the different elements in the system. This configuration includes some of the aspects defined in Section~\ref{sec:features}. For instance, defining whether (and how) the relying party will variate its QoS. It also includes configuration over the different reputation computation engines to determine which configuration adapts better to the defined scenario.
 
To cover the rest of the aspects described in Section~\ref{sec:features}, and taking into account other different reputation threats~\cite{trt:2009:gomez-marmol:cose}, we have defined and included in the simulator the following types of users, OpenID providers and relying parties acting in the system.

\begin{itemize}

\item Type of users:
\begin{itemize}
\item \textbf{Normal}: These users provide appropriate
recommendations according to the relying party quality of service.

\item \textbf{Negative or Positive raters}: These users always provides bad (negative raters) or good (positive raters) recommendations when giving feedbacks, regardless of the quality of the received service.


\end{itemize}

\item Type of OpenID providers:
\begin{itemize}
\item \textbf{Normal raters}: These OpenID providers properly follows the reputation framework definition. That is, they do not try to cheat when they provide recommendations.

\item \textbf{Negative or Positive raters}: These OpenID providers
always give bad (negative raters) or good (positive raters) recommendations about the relying party, regardless of its real behaviour, trying to decrease or increase its global reputation.


\item \textbf{Camouflaged positive/negative raters}: This kind of OpenID
provider is an extension of the previous one. It gives good or bad
recommendation, regardless of the real quality of service, but only
$p$\% of the times. The rest of times, $(100-p)$\%, it has a
normal behaviour.

\item \textbf{Sybil positive/negative raters}: These OpenID
providers act as positive or negative raters, but additionally, after a while, they replace its identity with a new one. 
\end{itemize}

\item Type of relying party:
\begin{itemize}
\item \textbf{Normal}: The relying party properly follows the behaviour of the framework without trying to cheat its reputation.

\item \textbf{Malicious}: The relying party includes in the recommenders list only the ones with better recommendations about itself.

\item \textbf{Sybil}: The relying party is disconnected and replaced with a new identity from time to time, reinitializing its associated reputation.

\item \textbf{Not participative}: The relying party does not
return the recommender list.
\end{itemize}

\end{itemize}

Once defined the scenario, the simulator starts the simulation environment. The simulation consists of executing a determined number of iterations. In each iteration, some of the simulated users want to access a service offered by the relying party, so they access the OpenID provider they belong to and ask for the reputation value of the relying party. Then, the OpenID provider collects recommendations using the technique described in Section~\ref{sec:background} and computes a reputation score according the weight given to each recommender.

If the relying party has enough reputation, the user will interact with the relying party and will provide a feedback according to the quality of the received service, and depending on the type of user. Finally, the OpenID provider adjusts the weights of its recommendation according to the received feedback.

\subsection{Visualization of results}

After running a simulation, the simulator shows three different charts, representing three different ways of analyzing the results. These charts are described in the following.

\begin{itemize}

\item \textbf{Results chart}. The Results chart compares the real Relying Party QoS in each moment with the average reputation value given by the Normal OpenID Providers. It is useful to see the behaviour of the reputation model against a specific scenario. For instance, whether it is resilient against malicious users or OpenID Providers or how long does it take to calculate the real reputation value.

\item \textbf{Accuracy chart}. Taking into account that the users interact with the relying party with a probability $p$, being $p$ the reputation given by its OpenID Provider, this chart determines how many users interact with a given relying party. This chart
is useful to compare different reputation computation engines regarding their accuracy when calculating reputation scores.


\item \textbf{User Satisfaction chart}. It indicates how satisfied the users are with the recommendations given by the OpenID Providers, and hence with the reputation framework.
This chart
is useful to compare different computation engines regarding the adaptation to users preferences. In other words, the more the computed reputation scores fit to each user, according to her preferences, the higher will be the user's satisfaction.


\end{itemize}

\section{Conclusions}\label{sec:conclusions}

Nowadays, a lot of service providers are including the OpenID standard solution as part of the authentication and basic access control services provided to their users. However, due to its decentralized nature, OpenID technology does not rely on any central authority which validates the trust level of the entities involved in the system, becoming a point where malicious service providers could try to gain access to the private attributes of users and make profit with them.

Some research has been done describing solutions helping to mitigate this problem. However, proposed solutions for this specific case still need some deep analysis and validation. In this paper we have presented a set of relevant features which should be taken into consideration to evaluate a reputation framework build on top of the OpenID standard. Based on this features, we have described a simulation environment able to validate the feasibility of such reputation frameworks and analyze their behaviour within different scenarios.

\bibliographystyle{splncs}
\bibliography{biblio}

\end{document}